\documentclass[prl,superscriptaddress,twocolumn,amsfonts,amsmath,a4paper,showpacs]{revtex4}
\usepackage{graphicx}

\begin{document}

\title{Anomalous Dynamic Arrest in a Mixture of Big and Small Particles}

%
%
\author{Angel J. Moreno}
\email[Corresponding author: ]{wabmosea@sq.ehu.es}
\affiliation{Donostia International Physics Center, Paseo Manuel de Lardizabal 4,
20018 San Sebasti\'{a}n, Spain.}
\author{Juan Colmenero}
\affiliation{Donostia International Physics Center, Paseo Manuel de Lardizabal 4,
20018 San Sebasti\'{a}n, Spain.}
\affiliation{Dpto. de F\'{\i}sica de Materiales, Universidad del Pa\'{\i}s Vasco (UPV/EHU),
Apdo. 1072, 20080 San Sebasti\'{a}n, Spain.}
\affiliation{Unidad de F\'{\i}sica de Materiales, Centro Mixto CSIC-UPV, 
Apdo. 1072, 20080 San Sebasti\'{a}n, Spain.}

\begin{abstract}

We present molecular dynamics simulations on the slow dynamics of a mixture
of big and small soft-spheres with a large size disparity. Dynamics are investigated
in a broad range of temperature and mixture composition. As a consequence of large size disparity,
big and small particles exhibit very different relaxation times.
As previously reported for simple models of short-ranged attractive colloids 
and polymer blends, several anomalous dynamic features are observed:
i) sublinear behavior for mean squared displacements, 
ii) concave-to-convex crossover for density-density correlators,
by varying temperature or wavevector,
iii) logarithmic decay for specific wavevectors of density-density correlators.
These anomalous features are observed over time intervals extending up to four decades,
and strongly resemble predictions of the Mode Coupling Theory (MCT)
for state points close to higher-order MCT transitions, which originate from the
competition between different mechanisms for dynamic arrest. 
For the big particles we suggest competition between soft-sphere repulsion
and depletion effects induced by neighboring small particles. For the small particles we suggest 
competition between bulk-like dynamics and confinement, respectively induced
by neighboring small particles and by the slow matrix of big particles.
By increasing the size disparity, a new relaxation scenario arises for the small particles.
Self-correlators decay to zero at temperatures where density-density correlations are frozen.
The behavior of the latters resembles features characteristic of type-A MCT transitions, defined by
a zero value of the critical non-ergodicity parameter.

\end{abstract}
\date{\today}
\maketitle

\begin{center}
\bf{I. INTRODUCTION}
\end{center}

Rheological properties of soft matter based systems can be manipulated by a proper addition
of components of different mobilities. Some examples are colloid-polymer mixtures and
polymer blends. These type of binary mixtures exhibit unusual relaxation features
which challenge standard pictures for structural dynamic arrest 
in glass-forming liquids and colloidal systems.
Differently from the usual two-step increase and decay for, respectively,
mean squared displacements and dynamic correlators \cite{mctrev1,mctrev2,koband,bennemann,aichele}, 
the latters do not exhibit a defined
plateau at intermediate times between the microscopic and diffusive regimes
\cite{prlsimA4,zaccarelli,blendpaper}.
This result suggests a softer character for the collective caging mechanism --- i.e., the temporary
trapping of each particle by its neighbors. 
Dynamic correlators show a highly stretched decay, and for selected values of the control
parameters the decay is logarithmic in time. By varying wavevectors or control parameters
the decay shows a striking concave-to-convex crossover \cite{prlsimA4,blendpaper}, leading to violation of 
standard scaling laws for complex dynamics as the time-temperature superposition principle.

These anomalous relaxation features have been recently derived within 
the Mode Coupling Theory (MCT) for simple models
of hard-sphere colloids with short-ranged attractions \cite{sperl}.
The MCT is a first-principle theory of the glass transition which makes predictions
for averaged dynamic quantities as mean squared displacements, diffusivities or incoherent
and coherent dynamic correlators \cite{mctrev1,mctrev2,das}.
The only input in MCT equations is the structural
information contained in total and partial static structure factors, which enter the memory function
accounting for the fluctuating forces. MCT determines {\it dynamic} quantities from 
the knowledge of {\it static} correlations. Indeed MCT relates small variations in static
correlations with large variations in dynamics. 

Solution of the MCT equations for the mentioned models of short-ranged 
attractive colloids \cite{sperl,fabbian,bergenholtz,dawson,jpcmA4} has determined the existence 
of so-called higher-order transitions, showing properties rather different from
the standard fold MCT transition associated to the liquid-glass transition.
The mentioned anomalous relaxation features have been related to the presence
of nearby higher-order transitions. MCT predictions for these type of models
have been confirmed by molecular dynamics simulations \cite{prlsimA4,zaccarelli,puertas}
and experiments \cite{mallamace,pham,eckert}.

The models of hard sphere-colloids with short-ranged attractions investigated
in Refs. \cite{sperl,prlsimA4,zaccarelli} are used as one-component effective models
{\it for the colloidal particles} in colloid-polymer mixtures. 
Addition of small polymers (or other small particles) to dense solutions of big colloidal particles
yields an effective attraction between the colloidal particles
in order to maximize entropy \cite{likos}.
This effect is known as the depletion mechanism. As a consequence, in a certain range of 
density, temperature and mixture composition, competition occurs between 
two different mechanisms for dynamic arrest of the colloidal particles: hard sphere-repulsion
characteristic of colloidal systems, and formation of reversible
bonds, induced by the depletion mechanism \cite{notegelnmax,gelnmax}. In the effective one-component system
the higher-order MCT scenario arises as a consequence of these two competing mechanisms
of very different localization lengths \cite{sperl,dawson}.
When heating up or cooling down the system, dynamic arrest is exclusively driven
by, respectively, hard-sphere repulsion and reversible bond formation, and relaxation features
of standard liquid-glass transitions are recovered \cite{prlsimA4,zaccarelli}.

Very recently, we have carried out simulations on a simple
bead-spring model for polymer blends \cite{blendpaper}.
The introduction of a signifficant monomer size disparity yields very different relaxation times
for both components in the blend, the component of small monomers being the fast one.
The fast component exhibits anomalous relaxation features, very different from
standard results observed for homopolymers, and strongly resembling predictions
of the higher-order MCT scenario for short-ranged attractive colloids. Fully atomistic simulations
on a real polymer blend are consistent with the anomalous features reported
for the bead-spring model \cite{genix}.
We have pointed out the hypothesis of an underlying higher-order MCT scenario for the dynamics
of the fast component, which might arise from the competiton between bulk-like caging and confinememt
as different mechanisms for dynamic arrest for the fast component. Bulk-like caging is induced by
the particles of the fast component, and confinement is induced by
the matrix formed by the chains of the slow component. Due to chain connectivity, the former 
mechanism is present even for high dilution of the fast component, 
extending the anomalous relaxation scenario
over a broad range of blend compositions \cite{blendpaper}.

At present no MCT theoretical calculations are available for models of polymer blends
similar to that investigated in Ref. \cite{blendpaper}.
A related system for which theoretical MCT works are available is a binary mixture
of big and small hard spheres \cite{gotzevoigtmann}. However, the size disparity
used in such works, and also in computational investigations in the MCT framework \cite{foffiprl,foffipre},
is not sufficient to provide a large separation in the time scales for big and small particles,
i.e., to induce confinement effects for the latters. It is not also sufficient to induce 
signifficant depletion effects for the big particles.
As a consequence, the anomalous relaxation features reported for polymer blends and 
short-ranged attractive colloids are not predicted for {\it moderate} size disparity, 
and indeed are not observed in the corresponding simulations. 
Instead, a description in terms of standard MCT predictions is possible for moderate size disparities
\cite{gotzevoigtmann,foffiprl,foffipre}.

Very recently, MCT theoretical calculations have been reported 
by Krakoviack \cite{krakoviackprl,krakoviackjpcm} for a binary mixture of {\it mobile and static} hard spheres,
a system where confinement effects are present \cite{notematrix,gallo}.
For this system, the extreme cases of the Lorentz gas 
(a single mobile particle in a disordered medium of static obstacles)
and the liquid of hard spheres are obtained for high dilution of respectively mobile and static particles.
At a given composition of the mixture a higher-order MCT transition has been derived.
This result supports the hypothesis of a similar MCT scenario in polymer blends
originating from competition between bulk-like caging and confinement.

The observed analogies between dynamics of colloidal particles in colloid-polymer mixtures, 
and of the fast component in polymer blends suggest that the higher-order MCT scenario
might be a general feature of systems showing slow dynamics 
with several competing mechanisms for dynamic arrest.
In this article we provide new evidence in favour of this hypothesis by
carrying out molecular dynamics simulations on a mixture of big and small
soft-spheres of very different sizes. Results presented here for large size disparity
(ratio $\delta = 2.5$) complement previous investigations on slow dynamics 
in binary mixtures with small disparity.
We observe anomalous relaxation features similar to those recently reported
for models of short-ranged attractive colloids and polymer blends. 
We investigate a wide range of temperatures and mixture compositions.
By tuning the composition, these features are displayed both by
the big and the small particles.

For the case of the big particles, we assign such anomalous features 
to competition between soft-sphere repulsion and the depletion mechanism
induced by the small particles. For the small particles, we suggest competition between 
bulk-like dynamics induced by the neighboring small particles, and confinement induced
by the matrix of slow big particles. Similarly to the fast component in polymer blends \cite{blendpaper},
and despite of the absence of chain connectivity, small
particles exhibit apparent anomalous relaxation over a broad range of compositions extending
up to high dilution. In the latter case, these effects are clearly manifested by a signifficant
subset of small particles forming small clusters.  

We have also performed simulations at a fixed composition for very large size disparity ($\delta = 8.0$).
A new relaxation scenario arises for the small particles,
showing features characteristic of nearby MCT transitions of the so called type-A. 
Such transitions are defined by a zero value of the long-time limit (non-ergodicity parameter)
of density-density correlators, different from the finite value defining the usual type-B transitions.
This feature provides a connection with results
in Ref. \cite{krakoviackprl} for a mixture of mobile and static particles,
which report a dynamic phase diagram displaying an A- and a B-line merging at a higher-order point.
Hence, we suggest that results at moderate disparity $\delta = 2.5$ for the system here investigated
might originate from the existence of a nearby B-line (providing finite values 
for the non-ergodicity parameters) ending at a nearby higher-order point 
(providing anomalous relaxation features).

The article is organized as follows. In Section II we introduce the investigated model and
give computational details. In Section III we present simulation results for static 
structure factors. We also present dynamic quantities displaying
unusual relaxation features. In Section IV the framework of the MCT is used 
in an operational way to describe simulation results. In Section V 
we discuss the possible origin of the observed anomalous dynamic features.
We also propose a picture for the different relaxation scenarios observed
by increasing the size disparity of the particles. Conclusions are given in Section VI.

\begin{center}
\bf{II. MODEL AND SIMULATION DETAILS}
\end{center}

We have simulated a mixture of big (labelled as A and B) and small (C and D) particles
of equal mass $m=1$, interacting through a soft-sphere potential plus a quadratic term:

\begin{equation}
V_{\alpha\beta} = 4\epsilon \left [\left (\frac{\sigma_{\alpha\beta}}{r}\right )^{12}
- C_0 + C_2\left (\frac{r}{\sigma_{\alpha\beta}}\right )^{2}\right ],
\label{eq:potsoft}
\end{equation}
where $\epsilon=1$ and $\alpha$, $\beta$ $\in$ \{A, B, C, D\}.
The interaction is zero beyond a cutoff distance $c\sigma_{\alpha\beta}$, with $c = 1.15$.
The addition of the quadratic term to the soft-sphere interaction, with
the values $C_0 = 7c^{-12}$ and $C_2 = 6c^{-14}$, guarantees continuity of potential and forces at the
cutoff distance.
The diameters of the soft-sphere potential for the different types of interaction are:
$\sigma_{\rm DD} =1$, $\sigma_{\rm CC} = 1.1\sigma_{\rm DD}$,  
$\sigma_{\rm BB} = 2.5\sigma_{\rm DD}$, $\sigma_{\rm AA} = 1.1\sigma_{\rm BB}$,
and $\sigma_{\alpha\beta}=(\sigma_{\alpha\alpha}+\sigma_{\beta\beta})/2$ for the case $\alpha \ne \beta$.

The potential (\ref{eq:potsoft}) is purely repulsive. It does not show local minima within 
the interaction range $r < c\sigma_{\alpha\beta}$. Hence, slow dynamics in the present model
arises as a consequence of steric effects. MCT theoretical works are usually carried out 
on systems of hard objects, while simulations in similar systems with continuous interactions
are usually preferred for computational simplicity.  In the present system, 
the tail of the interaction potential is progressively probed by decreasing temperature, 
which plays the role of increasing packing in a system of hard spheres. Hence, simulations
presented here should be useful for a {\it qualitative} test on the success or failure
of future MCT theoretical works on mixtures of hard spheres with large size disparity.

The composition of the mixture is defined as the fraction of small particles: 
$x_{\rm small} = (N_{\rm C} + N_{\rm D})/(N_{\rm A} + N_{\rm B} + N_{\rm C}+N_{\rm D} )$,
with $N_{\alpha}$ denoting the number of particles of the species $\alpha$.
As shown below, the introduction of a large size disparity  
between the sets \{A,B\} and \{C,D\} yields very different time scales for both sets.
We impose the constraints $N_{\rm A}=N_{\rm B}$ and $N_{\rm C}=N_{\rm D}$.
These  constraints, together with the small selected ratios $\sigma_{\rm CC}/\sigma_{\rm DD}=
\sigma_{\rm AA}/\sigma_{\rm BB} = 1.1$ guarantee that only very small dynamic differences 
are induced between particles within a same set (\{A,B\} or \{C,D\}), and at the same time,
avoid crystallization for the investigated compositions $x_{\rm small}$. 
Crystallization would occur for very asymmetric mixtures if only one type of big and small particles
were introduced \cite{notecrys}.

In the following, temperature $T$, distance, wavevector $q$, 
and time $t$, will be given respectively in units of $\epsilon/k_B$, $\sigma_{\rm DD}$,
$\sigma_{\rm DD}^{-1}$, and $\sigma_{\rm DD}(m/\epsilon)^{1/2}$.
The packing fraction, $\phi$, is defined as:

\begin{eqnarray} 
\nonumber \phi = \frac{\pi}{6L^3}[N_{\rm A}\sigma_{\rm AA}^3 + N_{\rm B}\sigma_{\rm BB}^3 \\
+ N_{\rm C}\sigma_{\rm CC}^3 + N_{\rm D}\sigma_{\rm DD}^3]
\label{eq:phi}
\end{eqnarray}
with $L$ the side of the simulation box. 
Simulations have been carried out at a constant packing fraction $\phi = 0.53$. 
This value is comparable to those used in simulations of slow relaxation in simple liquids.
For comparison, the original Lennard-Jones binary mixture investigated
by Kob and Andersen \cite{koband} has $\phi = 0.59$
with the definition of packing fraction given above.
We investigate the $T$-dependence of the dynamics for mixture compositions 
$x_{\rm small}=$ 0.1, 0.3, 0.6 and 0.8. The number of big and small particles 
for each composition are respectively ($N_{\rm A}+N_{\rm B}$:$N_{\rm C}+N_{\rm D}$)= 
(5400:600),  (2100:900), (1000:1500) and (800:3200).
The system is prepared by placing the particles
randomly in the simulation box, with a constraint that avoids core overlapping.
Periodic boundary conditions are implemented.
Equations of motion are integrated by using the velocity Verlet scheme \cite{frenkel},
with a time step ranging from $2 \times 10^{-4}$ to $5 \times 10^{-3}$,
for respectively the highest and the lowest investigated temperature.
A link-cell method \cite{frenkel} is used for saving computational time
in the determination of particles within the cutoff distance of a given one. 

At each state point, the system is thermalized at the requested temperature by periodic velocity rescaling.
After reaching equilibrium, energy and pressure show no drift.
Likewise, mean squared displacements and dynamic correlators show no aging, i.e., no time shift 
when being evaluated for progressively longer time origins. 
Once the system is equilibrated, a microcanonical run is performed for production
of configurations, from which static structure factors, mean squared displacements,
and dynamic correlators are computed. For each state point, the latter quantities are averaged over
typically 20-40 independent samples.

\begin{center}
\bf{III. RESULTS}
\end{center}

\begin{center}
\bf{a. Static structure factors}
\end{center}

\begin{figure}
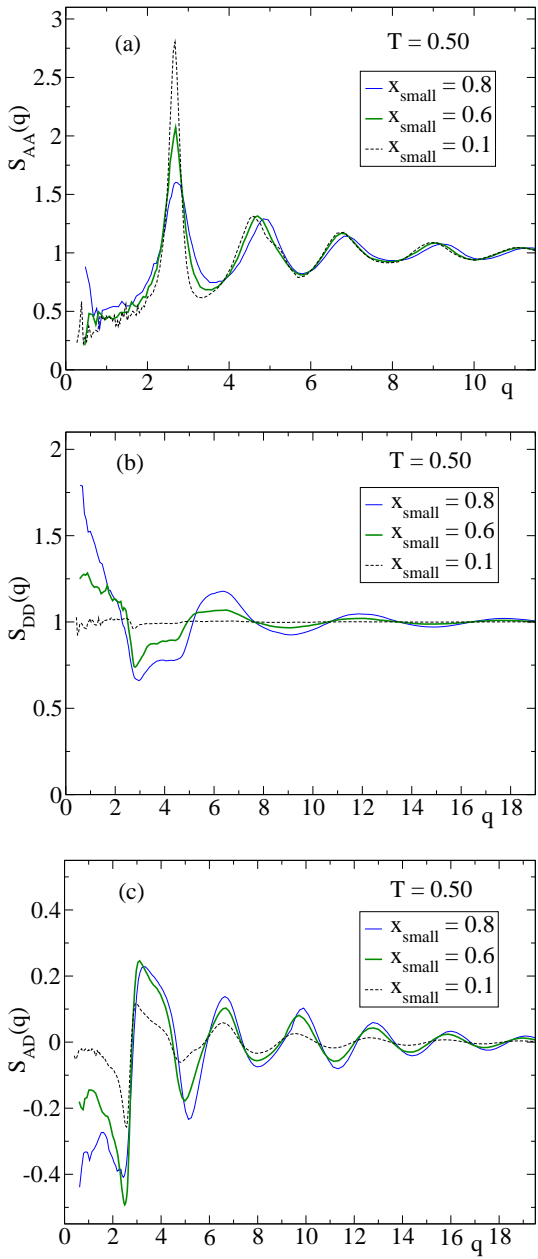

\includegraphics[width=0.86\linewidth]{fig1a.eps}
\newline
\newline
\includegraphics[width=0.86\linewidth]{fig1b.eps}
\newline
\newline
\includegraphics[width=0.86\linewidth]{fig1c.eps}
\newline
\caption{(color online) Partial static structure factors at $T=0.50$
and different mixture compositions, as computed from simulation data.
Wavevectors are given in units of $\sigma_{\rm DD}^{-1}$} 
\label{fig1}
\end{figure}

We compute normalized partial static structure factors, 
$S_{\rm \alpha\beta}(q) = \langle \rho_{\alpha}(q,0)\rho_{\beta}(-q,0)\rangle / \sqrt{N_{\alpha}N_{\beta}}$,
with $\rho_{\alpha}(q,t) = \Sigma_{j}\exp[i{\bf q}\cdot {\bf r}_{\alpha,j}(t)]$, 
the sum extending over all the particles of the species $\alpha \in$ \{A,B,C,D\}.
Fig. \ref{fig1} shows, at a fixed temperature $T = 0.50$ 
and different mixture compositions, results for A-A, D-D, and A-D pairs.
Data for other big-big, small-small, and big-small pairs display only small quantitative
differences. 

For high concentration of the big particles, $x_{\rm small}=0.1$, $S_{\rm AA}(q)$ shows
a sharp first peak at $q = 2.65\sigma_{\rm DD}^{-1}$, corresponding to a typical distance
of $2.37\sigma_{\rm DD}$ between neighboring A-particles. This distance is smaller
than the soft-sphere diameter $\sigma_{\rm AA} = 2.75\sigma_{\rm DD}$. This feature
is possible due to the interpenetrable character of soft spheres.
By increasing the concentration of small particles, the matrix of big particles 
progressively becomes more disordered. As a consequence, the height of the first peak 
of $S_{\rm AA}(q)$ decreases considerably. The first minimum follows the opposite trend 
and becomes less pronounced. The small shift of maxima and minima to higher $q$-values by increasing 
$x_{\rm small}$ reflects a stronger packing of the big particles.
For large concentration of small particles, $x_{\rm small}=0.8$,  $S_{\rm AA}(q)$
shows a small peak at low-$q$. We assign this peak to the presence of inhomogeneities or ``voids'' in
the matrix of big particles, which are filled by the small particles (see also Section V, Fig. \ref{fig11}c).

At high dilution of the small particles, $x_{\rm small}=0.1$,  $S_{\rm DD}(q)$
shows a nearly structureless profile, close to the flat behavior expected for a gas. 
Only a weak oscillation is observed at low-$q$. This feature reflects the existence of
a small fraction of clusters of neighboring small particles (see also Section V).
Increasing the concentration of small particles yields a broad peak at $q \approx 6.3\sigma_{\rm DD}^{-1}$,
corresponding to a typical distance of about $\sigma_{\rm DD}$ between neighboring small particles.
The peak grows up and narrows by increasing $x_{\rm small}$, as a signature of progressive ordering of the
small particles. Still, one finds a rather broad peak at the highest value of $x_{\rm small}=0.8$.
Increasing the concentration of small particles also produces a low-$q$ peak of increasing intensity
in $S_{\rm DD}(q)$. Such a peak originates from the inhomogeneities in the structure formed by the small
particles (see also Section V, Figs. \ref{fig11}b and \ref{fig11}c). We do not observe changes,
at any of the investigated temperatures, in the intensity of the low-$q$ peak within the time window
of the simulation. Hence, phase separation is discarded for the results here presented.

\begin{center}
\bf{b. Diffusivities and mean squared displacements}
\end{center}

\begin{figure}
\includegraphics[width=0.99\linewidth]{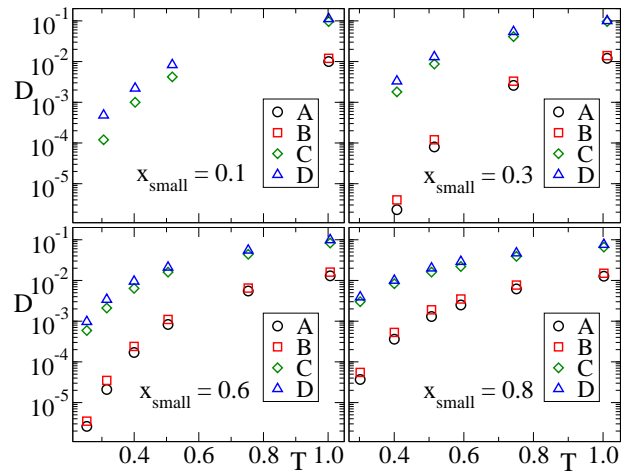}
\caption{(color online) For the investigated compositions, temperature dependence of the diffusivities
for the different species of the mixture. For low temperatures at $x_{\rm small}=0.1$, data
for the big particles are absent, since they do not reach the diffusive regime in the time window
of the simulation (see text and Fig. \ref{fig3}a).} 
\label{fig2}
\end{figure}

Fig. \ref{fig2} shows the $T$-dependence of the diffusivity, $D$, for all the species in the mixture
at all the investigated compositions. For each species $\alpha \in$ \{A,B,C,D\} the diffusivity
is calculated from the long time limit of $\langle [\Delta r_{\alpha}(t)]^2 \rangle/6t$,
with $\langle [\Delta r_{\alpha}(t)]^2 \rangle$ the corresponding mean squared displacement at time $t$.
As shown in Fig. \ref{fig2}, the introduction of a signifficant size disparity yields very different
time scales for the sets of big and small particles. Small differences are instead obtained 
between the diffusivities of both species within a same set.
Only for the lowest temperature at concentration $x_{\rm small} = 0.1$ there are signifficant differences 
(about a factor 4 in diffusivity) between C- and D-particles. 
In the following, simulation results will only be shown
for the big A-particles and the small D-ones.  The qualitive behavior of respectively B- and C-particles
is the same, displaying only small quantitative differences with the formers.

\begin{figure}
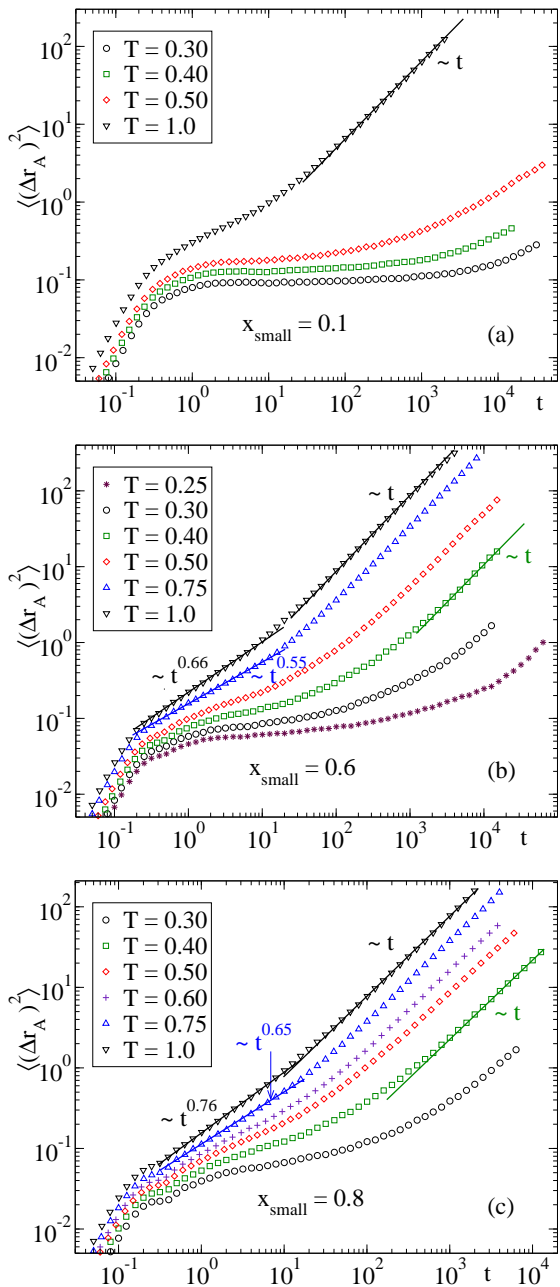

\includegraphics[width=0.895\linewidth]{fig3a.eps}
\newline
\newline
\includegraphics[width=0.895\linewidth]{fig3b.eps}
\newline
\newline
\includegraphics[width=0.895\linewidth]{fig3c.eps}
\newline
\caption{(color online) Symbols: mean squared displacement of A-particles for 
different compositions of the mixture. Straight lines indicate linear or sublinear behavior.} 
\label{fig3}
\end{figure}

Fig. \ref{fig3} shows the $T$-dependence of the mean squared displacement of the A-particles,
$\langle (\Delta r_{\rm A})^2 \rangle$, for three different compositions $x_{\rm small} =$ 0.1, 0.6 and 0.8.
As usually observed in the proximity of liquid-glass transitions \cite{koband,mctrev2,bennemann},
a bending occurs after the initial ballistic ($\propto t^2$) regime. 
A plateau arises at low temperatures. This effect
corresponds to the well-known caging regime ---i.e., the temporary trapping of each particle
in the cage formed by the neighboring ones. At long times, the diffusive regime ($ \propto t$) 
is reached for values $\langle (\Delta r_{\rm A})^2 \rangle \lesssim \sigma_{\rm AA}^2$, i.e, when
the A-particles have moved, on average, a distance of the order of their size.

\begin{figure}
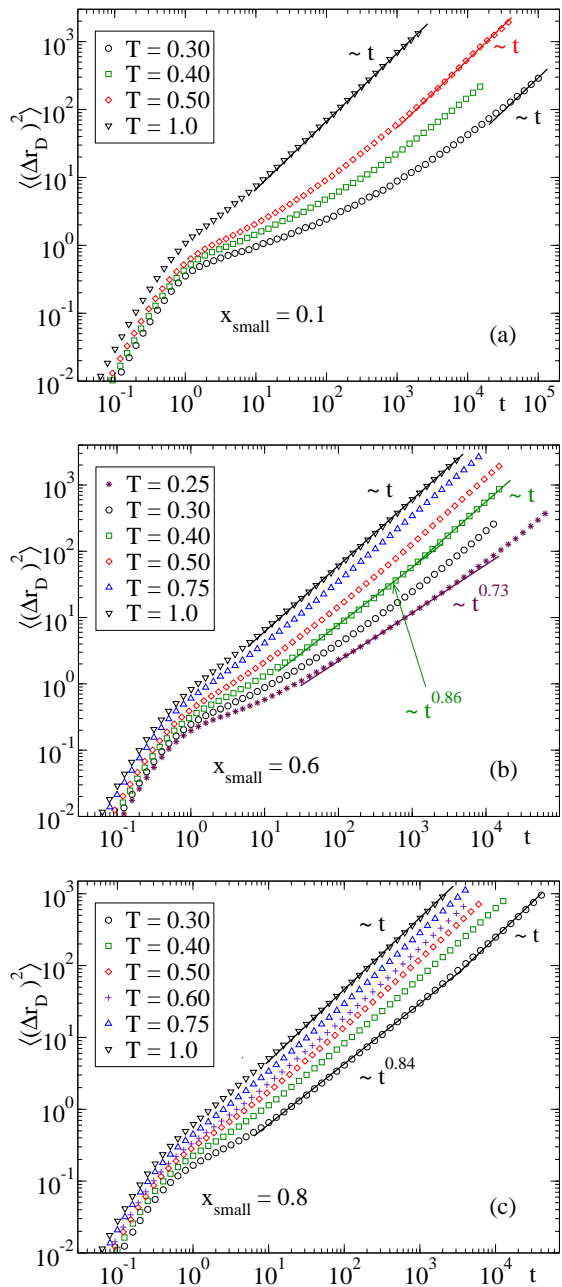

\includegraphics[width=0.895\linewidth]{fig4a.eps}
\newline
\newline
\includegraphics[width=0.895\linewidth]{fig4b.eps}
\newline
\newline
\includegraphics[width=0.895\linewidth]{fig4c.eps}
\newline
\caption{(color online) As Fig. \ref{fig3} for D-particles.} 
\label{fig4}
\end{figure}

By looking in more detail at the data for $T = 0.75$ and $T = 1.0$, at $x_{\rm small} \ge 0.6$
(Figs. \ref{fig3}b and \ref{fig3}c), an unusual appproximate sublinear behavior, $\propto t^\alpha$, 
is observed over two time decades after the ballistic regime. 
The exponent $\alpha < 1$ decreases by decreasing temperature.  
For $x_{\rm small} = 0.8$ we also note, at all temperatures, 
a bump at the interval $0.2 \lesssim t \lesssim 0.4$.

Fig. \ref{fig4} shows, at the same compositions, results for D-particles.
Remarkable differences with mean squared displacements of A-particles are observed.
Differently from A-particles (and from the standard behavior in the proximity of liquid-glass transitions)
D-particles reach the diffusive regime $\propto t$ for displacements much larger than their size: 
from  $\langle (\Delta r_{\rm D})^2 \rangle \sim 10\sigma_{\rm DD}^2$ at high $T$
to $\langle (\Delta r_{\rm D})^2 \rangle \sim 100\sigma_{\rm DD}^2$ at low $T$.
This result is observed for all the compositions.
As in the case of the A-particles, for $x_{\rm small} \ge 0.6$  an unusual approximate 
sublinear regime is observed at intermediate times, with an exponent decreasing
by decreasing temperature. This sublinear regime sets on
for $\langle (\Delta r_{\rm D})^2 \rangle \gtrsim \sigma_{\rm DD}^2$.

Results reported in this subsection evidence the existence of unusual relaxation features
in the slow dynamics of mixtures of big and small particles with sufficiently large size disparity. 
Next we evaluate the effects of size disparity in partial density-density correlators.

\begin{center}
\bf{c. Density-density correlators}
\end{center}

\begin{figure}
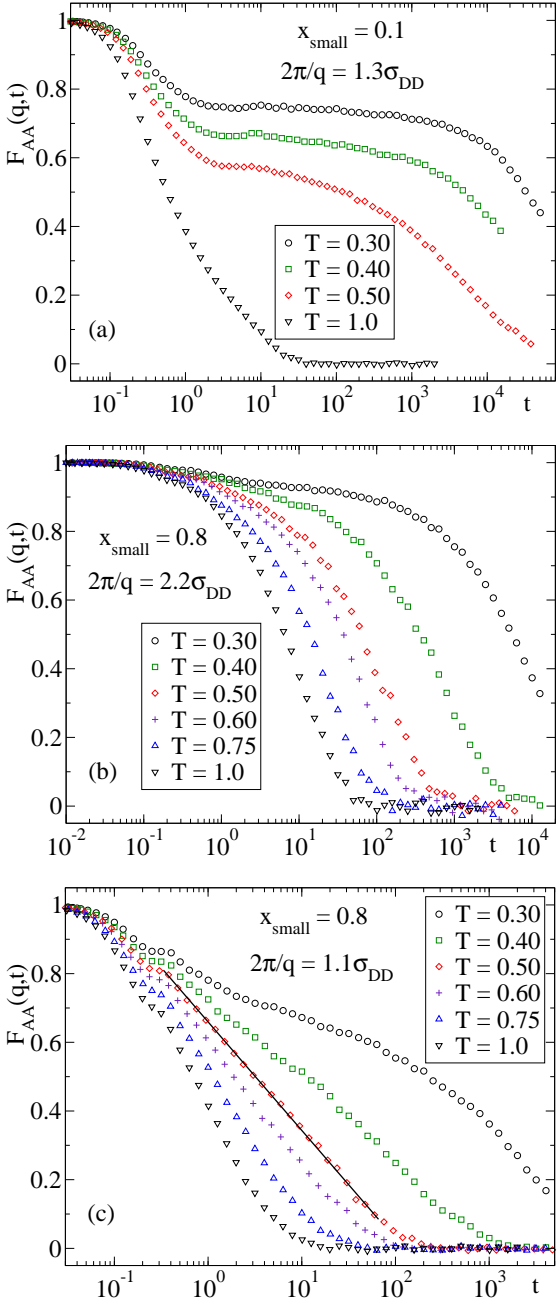

\includegraphics[width=0.895\linewidth]{fig5a.eps}
\newline
\newline
\includegraphics[width=0.895\linewidth]{fig5b.eps}
\newline
\newline
\includegraphics[width=0.895\linewidth]{fig5c.eps}
\newline
\caption{(color online) Symbols: $T$-dependence of $F_{\rm AA}(q,t)$, at specific $q$-values, 
for several compositions. The straight line in panel (c) indicates logarithmic decay.} 
\vspace{-2 mm}
\label{fig5}
\end{figure}

We compute partial density-density correlators,
$F_{\alpha\beta}(q,t)=\langle \rho_{\alpha}(q,t)\rho_{\beta}(-q,0)\rangle/
\langle \rho_{\alpha}(q,0)\rho_{\beta}(-q,0)\rangle$.
Fig. \ref{fig5} shows the $T$-dependence of $F_{\rm AA}(q,t)$ at specific values of the
wavevector $q$ and different compositions. 
For the case $x_{\rm small}=0.1$ (Fig. \ref{fig5}a), A-A correlations display
the standard behavior observed for liquid-glass transitions \cite{mctrev1,mctrev2,bennemann,aichele,koband}.
After the initial transient regime,
$F_{\rm AA}(q,t)$ shows a first decay to a plateau. By decreasing temperature,
the plateau extends over longer time intervals. At long times, a second decay
occurs from the plateau to zero. This second decay corresponds
to the $\alpha$-process of the glass transition and is usually well described by a stretched exponential function.

Figs. \ref{fig5}b and \ref{fig5}c show results for the case $x_{\rm small}=0.8$, for 
wavelengths ($2\pi/q$) of respectively $2.2\sigma_{\rm DD}$ and $1.1\sigma_{\rm DD}$.
While apparently standard behavior is obtained in the former case, unusual features are observed
for wavelengths probing the size of the small particles (Fig. \ref{fig5}c). 
First, $F_{\rm AA}(q,t)$ does not exhibit a defined plateau. Moreover, the shape of the long-time decay
shows a concave-to-convex crossover by decreasing temperature. At an intermediate temperature $T=0.50$
the decay is logarithmic over two time decades. As in the case of the mean squared displacement, 
a bump is observed at the interval $0.2 \lesssim t \lesssim 0.4$.

\begin{figure}
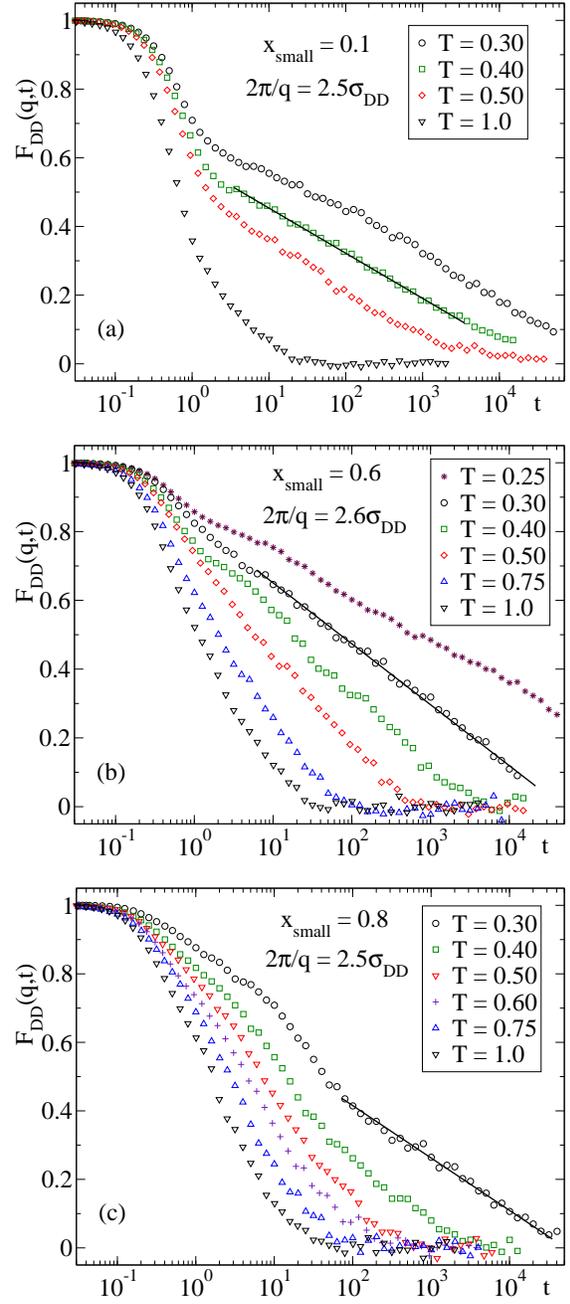

\includegraphics[width=0.895\linewidth]{fig6a.eps}
\newline
\newline
\includegraphics[width=0.895\linewidth]{fig6b.eps}
\newline
\newline
\includegraphics[width=0.895\linewidth]{fig6c.eps}
\newline
\caption{(color online) As Fig. \ref{fig5} for $F_{\rm DD}(q,t)$.} 
\vspace{-2 mm}
\label{fig6}
\end{figure}

Unusual relaxation features are also displayed by $F_{\rm DD}(q,t)$. Fig. \ref{fig6} shows
results for different compositions at wavelengths probing the size of the big particles.
No defined plateaux are observed for $x_{\rm small}=$ 0.1 and 0.6. Logarithmic relaxation
is observed at intermediate temperatures for time intervals extending up to three decades.
Data for $x_{\rm small}=0.6$ exhibit a concave-to-convex crossover.
Due to the mentioned partial crystallization of big particles for $x_{\rm small}=0.1$
\cite{notecrys}, data are absent in Fig. \ref{fig6}a for the range $0.50 < T < 1.0$. Apparently, a
concave-to-convex crossover is also present for this composition, though data are not conclusive. 

Results for $F_{\rm DD}(q,t)$ at $x_{\rm small}=0.8$ display a qualitatively different behavior.
At the lowest investigated temperature $T = 0.30$ a bump is observed at the interval $3 < t < 6$, 
followed by a decay until $t \approx 60$, where logarithmic relaxation sets
on and extends over three time decades.

\begin{center}
\bf{IV. MCT ANALYSIS}
\end{center}

\begin{center}
\bf{a) Big particles}
\end{center}

Many of the anomalous relaxation features reported in Section III --- sublinear behavior for mean squared
displacements, and logarithmic decay and concave-to-convex crossover for density-density correlators ---
strongly resemble those reported for hard-sphere colloids with short-ranged attractions \cite{prlsimA4,zaccarelli},
and for the fast component in polymer blends with components of very different mobilities \cite{blendpaper,genix}.
As mentioned in the Introduction, these anomalous features have been strictly 
derived in the framework of the MCT for simple models of short-ranged attractive colloids \cite{sperl}.
According to MCT, this anomalous relaxation scenario arises
from an underlying higher-order transition. 
Motivated by this fact, we discuss the present results by using MCT in an operational way.

In its ideal version, which neglects activated hopping events, MCT predicts
a sharp transition from an ergodic liquid to a non-ergodic arrested state at a given value
$x_{\rm c}$ of the relevant control parameter $x$ (in practice density or temperature) 
\cite{mctrev1,mctrev2,das}.
When crossing the transition point $x=x_{\rm c}$ from the ergodic to the arrested state, 
the long-time limit of the density-density correlator for wavevector $q$, $F(q,t)$,
jumps from zero to a non-zero value, denoted as the critical non-ergodicity parameter, $f^{\rm c}_q$. 
Moving beyond the transition point
into the non-ergodic state yields a progressive increase of the non-ergodicity parameter,
$f_q > f^{\rm c}_q$. In the MCT formalism, the standard liquid-glass transition is of the {\it fold} type,   
also denoted as A$_2$ \cite{mctrev1,mctrev2,das,noteorder}. In the standard case the jump in $F(q,t)$ 
is discontinuous, i.e., the critical non-ergodicity parameter $f^{\rm c}_q$ takes a finite value.
MCT transitions with $f^{\rm c}_q > 0$ are also denoted as {\it type-B} transitions.
For ergodic states close to the transition point, the initial part of the $\alpha$-process
 --- i.e., the von Schweidler regime --- is approximated by a power law expansion \cite{mctrev1,mctrev2,das}:
\begin{equation}
F(q,t) \approx f^{\rm c}_q -h_q (t/\tau)^{b} + h_q^{(2)}(t/\tau)^{2b},
\label{eqvonsch}
\end{equation}
with $0 < b \le 1$. The prefactors $h_q$ and $h_q^{(2)}$ only depend on the wavevector $q$. 
The characteristic time $\tau$ only depends on the separation parameter $|x - x_{\rm c}|$
and is divergent at the transition point. 
Another important prediction of the MCT for state points close to fold transitions
is the so-called second universality or $x$-time superposition principle (with $x$ the
corresponding control parameter). According to this prediction, the final decay of 
$F(q,t)$ (i.e., the final part of the $\alpha$-process) is unvariant under scaling
by the $\alpha$-relaxation time $\tau_{\alpha}$. Hence, for two state points $x=x_1$ and $x=x_2$
close to the transition point the final decay of $F(q,t)$ fulfills the relation \cite{mctrev1,mctrev2,das}:
\begin{equation}
F_{x=x_1}\left ( q,\frac{t}{\tau_{\alpha}(x_1)} \right ) 
= F_{x=x_2}\left (q,\frac{t}{\tau_{\alpha}(x_2)}  \right ).
\label{secuniv} 
\end{equation}
The $\alpha$-relaxation time $\tau_{\alpha}$ is a time scale probing the $\alpha$-process.
In practice, it can be obtained from fitting the decay from the plateau to a stretched exponential
$A_q\exp[-(t/\tau_{\alpha})^\beta]$, with $A_q$ the plateau height and $0 < \beta <1$. 
It can also be defined as the time where $F(q,t)$ decays to some small value, e.g. 0.3  ($\tau_{0.3}$), 
provided it is well below the plateau.

Another prediction of the MCT for points close to an A$_2$-transition is the power-law decrease 
of the diffusivity to zero, $D \propto |x-x_{\rm c}|^\gamma$.
The exponent $\gamma$ is given by \cite{mctrev1,mctrev2,das}:
\begin{equation}
\gamma = \frac{1}{2a} + \frac{1}{2b},
\label{eqgamma}
\end{equation}
with $0 < a < 0.4$. Hence $\gamma \ge 1.75$. The critical exponents $a$, $b$ and $\gamma$
are related to the so-called exponent parameter $1/2 \le \lambda \le 1$ through the relation:
\begin{equation}
\lambda = \frac{\Gamma^{2}(1+b)}{\Gamma(1+2b)} = \frac{\Gamma^{2}(1-a)}{\Gamma(1-2a)},
\label{eqlambda}
\end{equation}
with $\Gamma$ the Gamma function \cite{mctrev1,mctrev2,das}.

By numerically solving the MCT equations  for dynamic correlators,
transition points are determined as those where the respective long time limit exhibits
a jump from a zero to a non-zero value. 
From the knowledge of the total and partial static correlations
at the transition point, all the critical exponents and the coefficients 
in Eqs. (\ref{eqvonsch}, \ref{eqgamma}, \ref{eqlambda}) are univoquely determined \cite{mctrev1,mctrev2,das}. 
Solving the MCT equations is a difficult task, which in general is only possible
for rather simplified models of real systems.
Hence, instead of solving the equations, critical exponents and prefactors are often obtained
as fit parameters from simulation or experimental data. 
Consistency of the data analysis requires that the so-obtained critical exponents fulfill
Eqs. (\ref{eqgamma}) and (\ref{eqlambda}). 

\begin{figure}
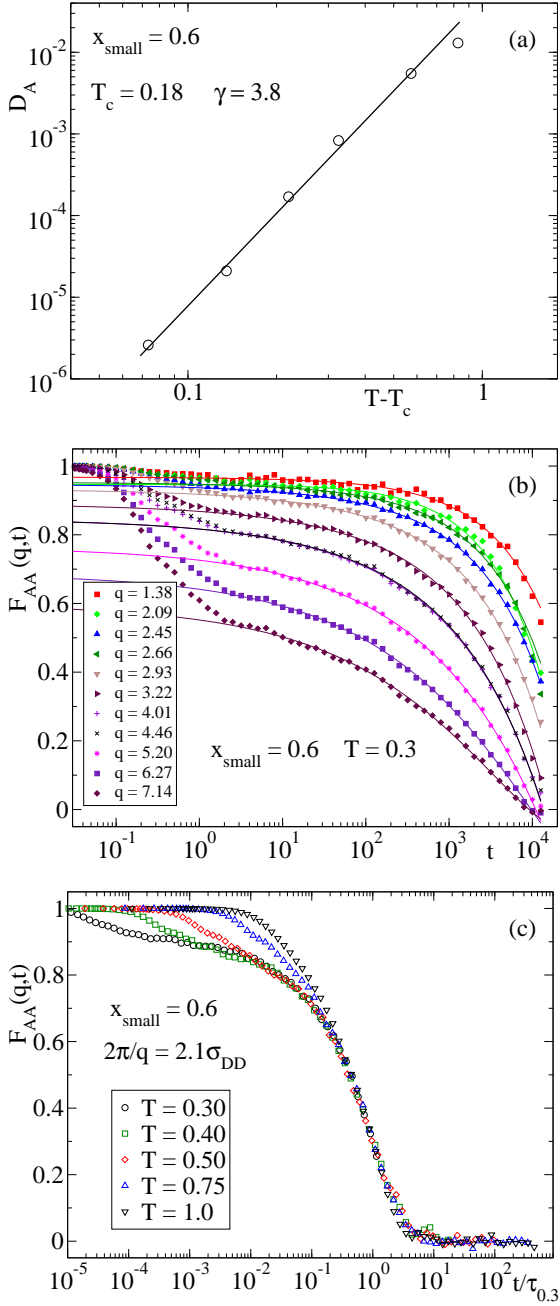

\includegraphics[width=0.895\linewidth]{fig7a.eps}
\newline
\newline
\includegraphics[width=0.895\linewidth]{fig7b.eps}
\newline
\newline
\includegraphics[width=0.895\linewidth]{fig7c.eps}
\newline
\caption{(color online) MCT analysis of data for A-particles, for composition $x_{\rm small}=0.6$.
Panel (a): $T$-dependence of the diffusivity. Symbols are simulation data. 
The line is a fit to a MCT power law $D_{\rm A} \propto (T-T_{\rm c})^\gamma$ (see text). 
Panel (b): Symbols: $q$-dependence  of $F_{\rm AA}(q,t)$  at a low temperature.
Curves are fits to Eq. (\ref{eqvonsch}). 
Panel (c): $T$-dependence of $F_{\rm AA}(q,t)$ for a specific $q$-value. 
In order to test the second universality, times are divided
by the $\alpha$-relaxation time $\tau_{0.3}(T)$ (see text).}
\label{fig7}
\end{figure}

Fig. \ref{fig7} shows an analyisis of data for A-particles at composition $x_{\rm small}=0.6$
in the framework of the MCT for A$_2$-transitions. Fig. \ref{fig7}a shows 
a fit of the diffusivity to a power law $D_{\rm A} \propto (T-T_{\rm c})^\gamma$.
The fit provides the values $T_{\rm c}=0.180$ and $\gamma = 3.80$.
From Eqs. (\ref{eqgamma}) and (\ref{eqlambda}) we obtain $\lambda=0.887$, $a=0.220$, and $b=0.326$.
Fig. \ref{fig7}b shows, for $F_{\rm AA}(q,t)$ at a low $T$, fits of the decay from the plateau
to Eq. (\ref{eqvonsch}). By forcing the von Schweidler exponent to the mentioned value $b=0.326$
a consistent description is achieved. Fig. \ref{fig7}c shows a succesful test of the second
universality (time-temperature superposition principle), by using $\tau_{0.3}$ 
for the $\alpha$-relaxation time. For the composition $x_{\rm small}=0.3$ similar 
results are obtained (not shown). From an analogous analysis we obtain 
$T_{\rm c} = 0.344$, $\lambda = 0.880$, $\gamma = 3.68$, $a = 0.226$, and $b = 0.340$
for this latter composition.

For the composition $x_{\rm small}=0.1$, a test of MCT for A-particles is not possible due
to the absence of data in the range $0.50 < T < 1.0$. Though a two-step decay is observed
for $T \le 0.50$ (see Fig. \ref{fig5}a), a fit to Eq. (\ref{eqvonsch}) is not possible in that 
temperature range, since the plateau height is not constant, but clearly increases by decreasing temperature.
According to predictions of {\it ideal} MCT, these temperatures would correspond to non-ergodic states
below the transition point ($T < T_{\rm c}$).
Within this interpretation, the fact that a decay from the plateau occurs at long times,
would be a signature of the so-called hopping events, which restore ergodicity.  

\begin{figure}
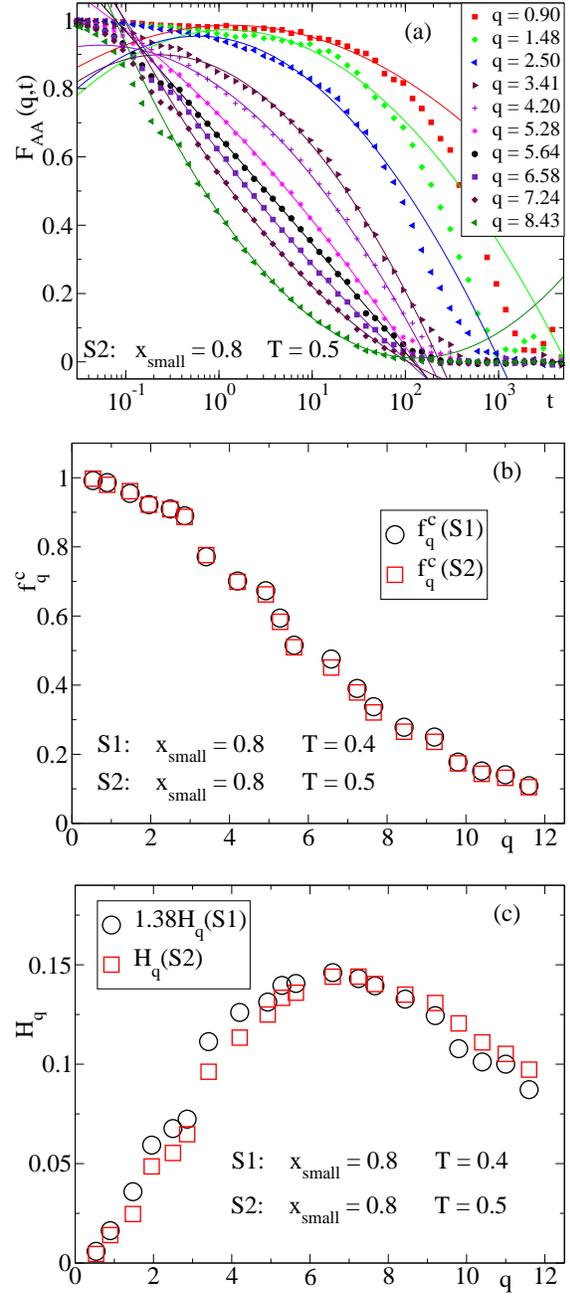

\includegraphics[width=0.895\linewidth]{fig8a.eps}
\newline
\newline
\includegraphics[width=0.895\linewidth]{fig8b.eps}
\newline
\newline
\includegraphics[width=0.895\linewidth]{fig8c.eps}
\newline
\caption{(color online) Panel (a): Symbols: $q$-dependence of $F_{\rm AA}(q,t)$ for 
a composition $x_{\rm small}=0.8$ at a low temperature.
Curves are fits to Eq. (\ref{eqlog}).
Panels (b) and (c): Values of, respectively, $f^{\rm c}_{q}$  and $H_{q}$,
as obtained from fits to Eq.~(\ref{eqlog}) of $F_{\rm AA}(q,t)$
at two state points S1 and S2. 
The characteristic times $\tau$ for S1 and S2 are respectively 9 and 3.}
\label{fig8}
\end{figure}

For the composition $x_{\rm small}=0.8$, data of $F_{\rm AA}(q,t)$ for wavelengths $2\pi/q$ probing
the size of the small particles do not show a defined plateau (Fig. \ref{fig5}c), and a fit
to  Eq. (\ref{eqvonsch}) is not possible. Moreover, the observed logarithmic decay and concave-to-convex 
crossover do not fit to expectations for A$_2$-transitions. They instead resemble
the features reported for short-ranged attractive colloids \cite{prlsimA4,zaccarelli}
and polymer blends \cite{blendpaper}, which for the former
case have been derived in the framework of MCT for state points close to higher-order transitions.

A higher-order MCT transition, A$_{n+1}$, is characterized 
by an exponent parameter $\lambda =1$, and can emerge as the result
from the interplay between $n$ control parameters $\{{\bf x}^n\} =$  $( x_1, x_2 ... x_n )$.
Higher-order MCT transitions were initially derived for schematic models \cite{schematic1,schematic2}, 
but only recently have been obtained for short-ranged attractive colloids 
\cite{bergenholtz,dawson,sperl,fabbian,jpcmA4}
as a first realization in real systems. It can be demonstrated that
an expansion in a power-law series as Eq. (\ref{eqvonsch}) is not convergent
for $\lambda = 1$ or for values of $\lambda$ very close to unity. On the contrary,
a logarithmic expansion is rapidly convergent. Hence, close to a higher-order MCT transition, 
or more generally to an A$_2$-transition with $\lambda \lesssim 1$, $F(q,t)$ is
approximated at intermediate times as \cite{schematic1,schematic2,sperl,notefqc}: 
\begin{equation}
F(q,t) \approx f^{\rm c}_q -H_q \ln (t/\tau) + H_q^{(2)}\ln^2 (t/\tau),
\label{eqlog}
\end{equation}
where the prefactors $H_q$ and $H_q^{(2)}$ depend on $q$ and on the distance
of the state point $\{{\bf x}^n\}$ to the transition point $\{{\bf x}^n_c\}$.
These prefactors exhibit two important properties \cite{sperl}: i) $H_q$ factorizes
as $H_q = C(\{{\bf x}^n\}) \tilde{H}_q$ where $\tilde{H}_q$ 
only depends on $q$, and the $q$-independent term $C(\{{\bf x}^n\})$ depends on the state point. 
Hence, values of $H_q$ for different state points close to the transition point
must be proportional. ii) $H_q^{(2)}$ does not follow scaling behavior. 
It is decomposed as \cite{sperl}:
\begin{equation} 
H_q^{(2)}=A(\{{\bf x}^n\}) + B(\{{\bf x}^n\})K_q , 
\label{eqhq2}
\end{equation}
where $K_q$ only depends on $q$, and the $q$-independent terms $A(\{{\bf x}^n\})$ and $B(\{{\bf x}^n\})$
depend on the state point. The terms $\tilde{H}_q$ and $K_q$ are univoquely determined by 
static structure factors at the higher-order transition point.
As in the case of A$_2$-transitions, coefficients in Eq. (\ref{eqlog}) 
are often obtained as fit parameters from simulation or experimental data.

Decomposition of $H^{(2)}_q$ according to Eq. (\ref{eqhq2}) has an important consequence.
There are hypersurfaces in the control parameter space, $\{{\bf x}^n\}$ = $\{{\bf x}^n\}(q)$, 
where $H^{(2)}_q$ changes its sign, being zero along the hypersurface. This property leads, for a given value of $q$,
to a concave-to-convex crossover in $F(q,t)$ when crossing the hypersurface by 
varying control parameters (as temperature or density) \cite{sperl}. Analogously, for a given state point, varying the value
of $q$ also leads to a concave-to-convex crossover in $F(q,t)$. Since $H^{(2)}_q = 0$ for state points 
at the hypersurface, according to Eq. (\ref{eqlog}) $F(q,t)$ will exhibit a logarithmic decay for such state points. 
Moving between different state points changes the value of $q$ for which pure logarithmic decay occurs.
The concave-to-convex crossover is one of the main signatures of the higher-order MCT scenario
and differentiates it from other theoretical frameworks \cite{sperl}.

As shown in Figs. \ref{fig5}c and \ref{fig8}a, a concave-to-convex crossover is present,
both by varying temperature and wavevector, for $F_{\rm AA}(q,t)$ at composition $x_{\rm small}=0.8$.
Fig. \ref{fig8}a shows fits to Eq. (\ref{eqlog}) for $T=0.50$ \cite{notelowq}. Figs. \ref{fig8}b and \ref{fig8}c
show respectively the obtained values for $f^{\rm c}_q$ and $H_q$, for temperatures $T=0.40$ and $T=0.50$.
The fact that a common $f^{\rm c}_q$ is found for both states, 
together with the observed scaling behavior of $H_q$, would be consistent with the existence
of a nearby MCT higher-order transition ($\lambda=1$) or an A$_2$-transition 
with $\lambda \lesssim 1$.  Differently from the usual behavior for liquid-glass transitions, 
$f^{\rm c}_q$ shows no strong modulation but a nearly monotonous decay. Though this result 
qualitatively resembles the observed $q$-dependence at A$_3$- and A$_4$- transitions for models
of short-ranged attractive colloids \cite{sperl,dawson,jpcmA4,prlsimA4,notemix},
such a comparison can be misleading. The observed $q$-dependence is not necessarily related to
a hypothetical higher-order MCT transition. It is indeed also found at 
moderate size disparity, for which standard relaxation features are observed \cite{gotzevoigtmann,foffipre}.

It is also noteworthy that the observed sublinear behavior in the mean squared displacement
at intermediate times is another feature characteristic of the higher-order MCT scenario.
In leading-order \cite{sperl}, the mean squared displacement
is given by $\langle [\Delta r(t)]^2 \rangle = r_{\rm c}^2 (t/\tau)^z$, with $r_{\rm c}$
the localization length and $z = h_{\rm MSD}C(\{{\bf x}^n\})/r_{\rm c}^2$.
The coefficient $h_{\rm MSD}$ is determined by static structure factors at the transition point.
Since the prefactor $C(\{{\bf x}^n\})$ decreases as the transition point is approached, the exponent $z$
also decreases, as observed in Fig. \ref{fig3}c. 
Deviations from pure power-law behavior result from corrections to the leading term.
Such corrections are minimal for certain values of the control parameters,
yielding pure power-law behavior \cite{sperl}.

\begin{center}
\bf{b) Small particles}
\end{center}

\begin{figure}
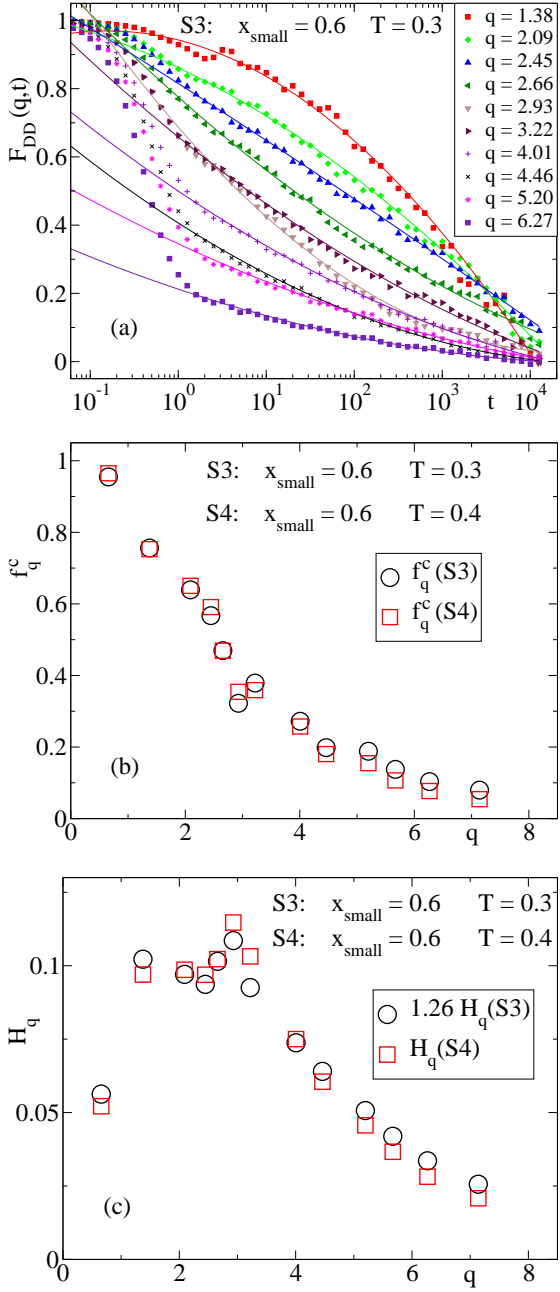

\includegraphics[width=0.895\linewidth]{fig9a.eps}
\newline
\newline
\includegraphics[width=0.895\linewidth]{fig9b.eps}
\newline
\newline
\includegraphics[width=0.895\linewidth]{fig9c.eps}
\newline
\caption{(color online) As Fig. \ref{fig8} for $F_{\rm DD}(q,t)$, $x_{\rm small} = 0.6$,
at the state points S3 and S4.
The characteristic times $\tau$ for S3 and S4 are respectively 30 and 7.2.
Data in panel (a) for the wavevector $q = 2.45\sigma_{\rm DD}^{-1}$ exhibit logarithmic decay
over four time decades.}
\label{fig9}
\end{figure}

Data for D-particles cannot be reproduced by the standard A$_2$-scenario for any of the investigated
compositions, due to the absence of defined plateaux and to the presence of logarithmic relaxation
in density-density correlators.
Data for $x_{\rm small}=0.6$ exhibit features 
resembling predictions of the higher-order MCT scenario. Fig. \ref{fig9} shows results
from such an analysis, analogous to that in Fig. \ref{fig8} for A-particles, 
in terms of the latter framework. The concave-to-convex crossover
in $F_{\rm DD}(q,t)$ obtained by decreasing temperature at constant $q$ (Fig. \ref{fig6}b),
is also observed by varying $q$ at constant temperature (Fig. \ref{fig9}a).
A good description of the decay is achieved by Eq. (\ref{eqlog}). Consistently,
fits for different state points provide the same values for $f_{q}^{\rm c}$
(Fig. \ref{fig9}b), and scaling behavior is obtained for the 
corresponding values of $H_q$ (Fig. \ref{fig9}c). The observed sublinear behavior for
the mean squared displacement, and the decrease of the exponent by decreasing temperature
(Fig. \ref{fig4}b) is another feature shared with the higher-order MCT scenario.

\begin{figure}
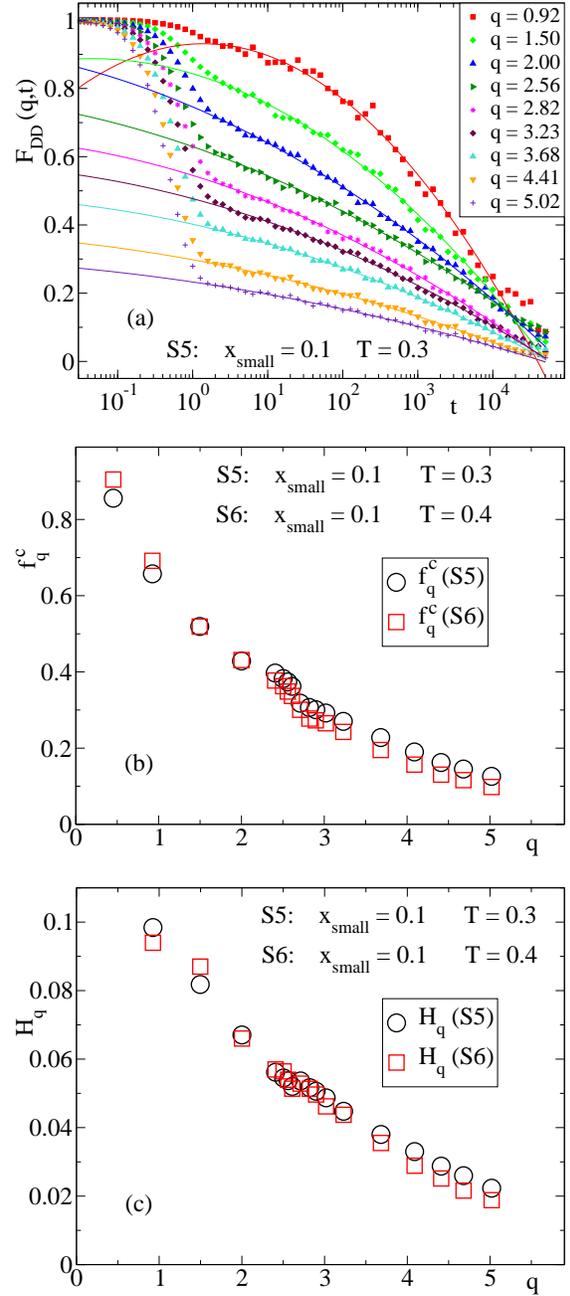

\includegraphics[width=0.895\linewidth]{fig10a.eps}
\newline
\newline
\includegraphics[width=0.895\linewidth]{fig10b.eps}
\newline
\newline
\includegraphics[width=0.895\linewidth]{fig10c.eps}
\newline
\caption{(color online) As Fig. \ref{fig8} for $F_{\rm DD}(q,t)$, $x_{\rm small} = 0.1$,
at the state points S5 and S6.
The characteristic times $\tau$ for S5 and S6 are respectively 360 and 50.}
\label{fig10}
\end{figure}

We have performed an analogous analysis for $F_{\rm DD}(q,t)$
at composition $x_{\rm small}=0.1$ (see Fig. \ref{fig10}),
despite there are some differences with features characteristic of the higher-order MCT scenario.
Differently from the case $x_{\rm small}=0.6$, no apparent sublinear behavior is present
in the mean squared displacement (Fig. \ref{fig4}a). 
Due to the absence of data in the range $0.5 < T < 1.0$, it is difficult to
unambiguously identify a concave-to-convex crossover in $F_{\rm DD}(q,t)$ 
by varying temperature (Fig. \ref{fig6}a).
Fig. \ref{fig10}a shows results by varying $q$ at constant $T=0.30$. The trend exhibited by the data,
which show a extremily stretched decay at large wavevectors,
suggests that such a crossover might be present for higher values of $q$.
However, we cannot confirm this point, which remains to be understood. 
We have not detected the crossover at least up to $q \approx 7\sigma_{\rm DD}^{-1}$.
Beyond that $q$-value the amplitude of the decay is rather small and 
it is difficult to solve the shape of the curve within statistical noise.

Still, Eq. (\ref{eqlog}) provides a good description of the decay (Fig. \ref{fig10}a), with a common
$f_q^{\rm c}$ for different state points (Fig. \ref{fig10}b), and $H_q$ obeying scaling behavior
(Fig. \ref{fig10}c). Surprisingly, the scaling factor is unity within error bars, despite 
dynamics at state points $T=0.30$ and $T=0.40$ are signifficantly different
(see Figs. \ref{fig4}a and \ref{fig6}a).

Due to the complex form of the decay of $F_{\rm DD}(q,t)$  for composition $x_{\rm small}=0.8$
(Fig. \ref{fig6}c), a fit to Eqs. (\ref{eqvonsch}) or (\ref{eqlog}) is not possible. 
Hence, at this composition there are no apparent analogies for small particles with known MCT scenarios.

\begin{center}
\bf{V. DISCUSSION}
\end{center}

Results presented in Sections III and IV exhibit strong dynamic analogies with 
MCT predictions for state points close to higher-order transitions
($\lambda = 1$), or more generally, for A$_2$-transitions with exponent 
parameter $\lambda \lesssim 1$. These analogies must not be understood as
a proof of an underlying MCT scenario for the mixture of big and small particles
here investigated. An unambiguous answer to this question could only be provided by solving
the corresponding MCT equations. As mentioned in the Introduction, theoretical and computational
works on the framework of the MCT on mixtures of big and small hard spheres have not reported
anomalous relaxation features \cite{gotzevoigtmann,foffiprl,foffipre}.
However these works have explored size disparities smaller than the
value $\delta = \sigma_{\rm AA}/\sigma_{\rm CC} = \sigma_{\rm BB}/\sigma_{\rm DD}= 2.5$ used in this work.

Fig. \ref{fig11} displays typical slabs of the simulation box for mixture compositions
$x_{\rm small} =$ 0.1, 0.6, and 0.8. The slab thickness is $5\sigma_{\rm DD}$.
On the basis of particular features displayed by the
configuration of small and big particles at the different compositions, next we discuss
the observed anomalous relaxation features in terms of competition between
different arrest mechanisms.

\begin{figure}
\includegraphics[width=0.78\linewidth]{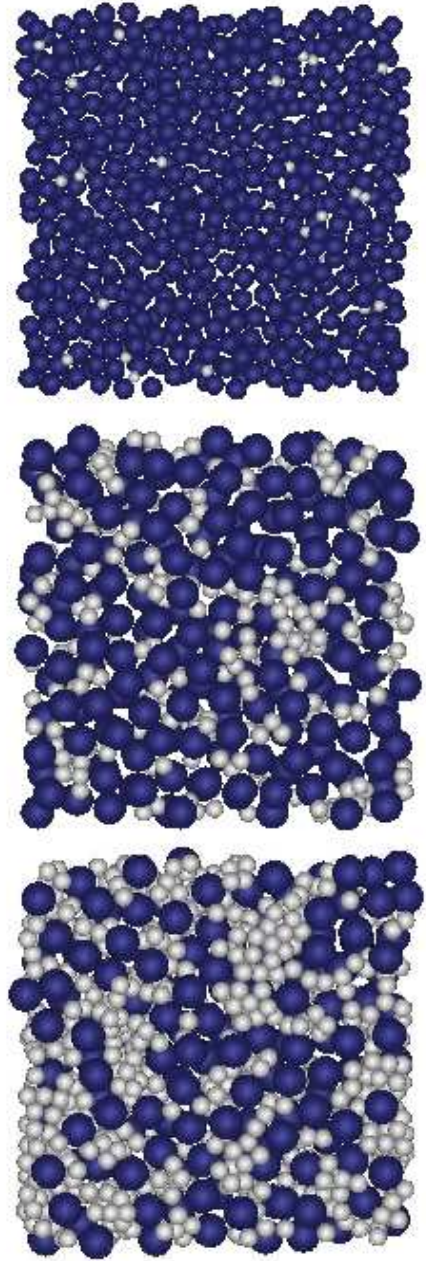}
\newline
\caption{(color online) Typical slabs (of thickness $5\sigma_{\rm DD}$) 
of different configurations at $T=0.30$.
From top to down $x_{\rm small}=$ 0.1, 0.6, and 0.8. 
Dark: A- and B-particles. Light: C- and D-particles.}
\label{fig11}
\end{figure}

\begin{center}
\bf{a) Symmetric mixtures}
\end{center}

For composition $x_{\rm small}=0.6$, big particles are 
distributed over the simulation box in a rather homogeneous way, with 
a weak tendency to form clusters. Small particles fill connected cavities in the slow matrix
of big particles, which acts as a confining medium for the formers.
As a consequence, competition between bulk-like dynamics and confinement occurs
for arrest of the small particles.
Pure logaritmic decay in $F_{\rm DD}(q,t)$ is observed 
for $q \sim 2.5\sigma_{\rm DD}^{-1}$ (Fig. \ref{fig9}a), i.e.,
for wavelengths $\sim 2.5\sigma_{\rm DD}$ probing the cavity size (Fig. \ref{fig11}, middle).

Dynamic arrest of big particles is essentially driven by soft-sphere repulsion, as 
it would be for a system without size disparity, and a consistent description of dynamic features
is achieved within the standard A$_2$-scenario of the MCT. Still, it must be mentioned
that the obtained value of the exponent parameter $\lambda = 0.887$ is 
signifficantly higher than typical values $\lambda \sim 0.7$ usually obtained within
the A$_2$-scenario \cite{mctrev2,bennemann,aichele,koband}. This high value of $\lambda$ can be interpreted 
as a signature of the depletion mechanism induced by neighboring small particles,
which leads to the observed weak clusterization of big particles.
Though only weakly competing with soft-sphere repulsion, the depletion mechanism
yields precursor effects of an incoming higher-order scenario ($\lambda \rightarrow 1$),
as strong stretching for the long-time decay in $F_{\rm AA}(q,t)$ (Fig \ref{fig7}b), 
or sublinear behavior in the mean squared displacement (Fig \ref{fig3}b).
These effects are weaker for the composition $x_{\rm small}=0.3$, yielding
a lower exponent parameter $\lambda = 0.880$.

\begin{center}
\bf{b) High concentration of small particles}
\end{center}

For composition $x_{\rm small}=0.8$, the population of small particles is sufficient 
to provide an efficient depletion mechanism, yielding a strong clusterization of the big particles
(Fig. \ref{fig11}, bottom). Competition
between soft-sphere repulsion and depletion leads to anomalous
relaxation features for the big particles. Hence, $F_{\rm AA}(q,t)$ shows logarithmic decay
for $q \sim 5.6\sigma_{\rm DD}^{-1}$ (Fig. \ref{fig8}a), i.e., for wavelengths 
$\sim \sigma_{\rm DD}$ probing the size of the small particles.

Though depletion effects are evidenced by clusterization of big particles, it is worth emphasizing
that a direct comparison with results for the effective one-component systems displaying
the higher-order MCT scenario \cite{sperl,jpcmA4,prlsimA4,zaccarelli,dawson,bergenholtz,fabbian}
cannot be made. In the effective one-component systems the small particles are absent and
the depletion mechanism is described by an effective short-ranged attraction
between the big particles \cite{likos}. However, concerning dynamics, 
the validity of this approximation does not only require a much smaller size, 
but also a much smaller mass for the small particles \cite{binreent}.  
In the case here investigated big and small particles have the same mass. 
Therefore, a proper proof of the existence of an underlying higher-order MCT scenario
can only be provided by solving the MCT equations for the two-component system. 
 
Density-density correlators for small particles
at $x_{\rm small} = 0.8$ exhibit a complex decay. An apparent logarithmic decay
occurs at long times (Fig. \ref{fig6}c). For $T = 0.30$ the onset of this decay occurs 
at $t \approx 60$, which corresponds to mean displacements 
$\langle (\Delta r_{\rm D})^2 \rangle^{1/2} \sim 2\sigma_{\rm DD}$ (Fig. \ref{fig4}c).
If logarithmic decay is again interpreted as the result of a competition between bulk-like dynamics
and confinement, the former distance would correspond to the length scale beyond which
confinement induced by the big particles affects dynamic arrest of the small particles.
For smaller length scales, the only relevant arrest mechanism is bulk-like caging induced 
by neighboring small particles. Hence, for  $t \lesssim 60$
small particles do not exhibit clear signatures of anomalous relaxation, as suggested
by Figs. \ref{fig4}c and \ref{fig6}c.

\begin{center}
\bf{c) Low concentration of small particles}
\end{center}

For $x_{\rm small}=0.1$, most of the small particles are isolated in the 
slow matrix of big particles (Fig. \ref{fig11}, top). 
However, there is also a signifficant population of pairs of neighboring small particles.
We have estimated that about a 7 \% of the small particles
have a neighboring small particle within a distance $1.3\sigma_{\rm DD}$. 
Clusters of three or more neighboring small particles for this inter-particle distance
are very rare. 

\begin{figure}
\includegraphics[width=0.8\linewidth]{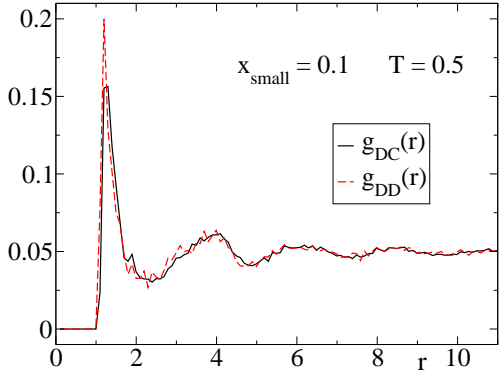}
\newline
\caption{(color online) Radial distribution function for D-C and D-D pairs, 
for $x_{\rm small} = 0.1$ and $T = 0.50$.}
\label{fig12}
\end{figure}
\begin{figure}
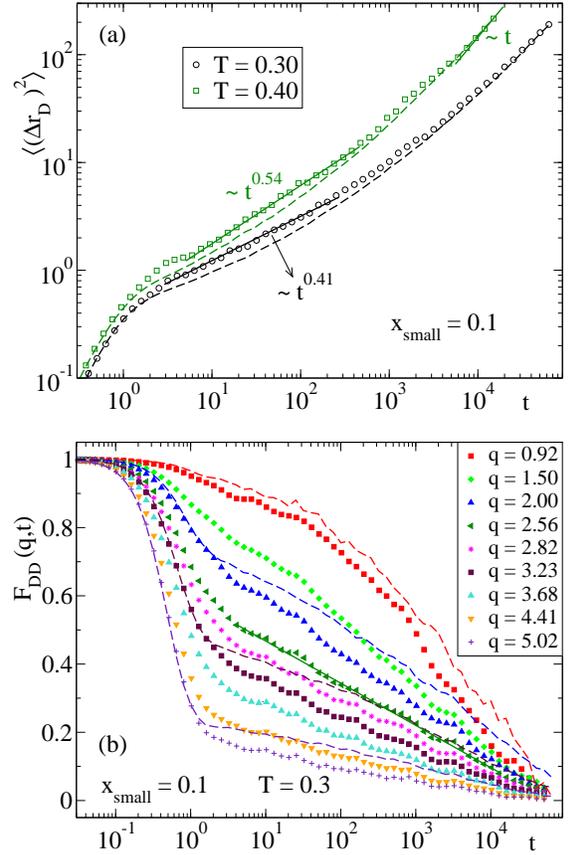

\includegraphics[width=0.895\linewidth]{fig13a.eps}
\newline
\newline
\includegraphics[width=0.895\linewidth]{fig13b.eps}
\newline
\caption{(color online) Symbols: For composition $x_{\rm small}=0.1$,
mean squared displacement (panel (a)), and
density-density correlator (panel (b)) for the subset of D-particles
which initially have at least one neighboring small particle 
within a distance $r < 1.3\sigma_{\rm DD}$.
Panel (a) shows data for $T=$ 0.30 and 0.40. Dashed curves are the corresponding data
by averaging over all the D-particles. Solid lines indicate linear or sublinear behavior.
Panel (b) shows data for different $q$-values at $T = 0.30$. Dashed curves are
data by averaging over all the D-particles (from top to down 
$q\sigma_{\rm DD}=$ 0.92, 2.00, 3.23, and 5.02).
The solid line indicates logarithmic relaxation.} 
\label{fig13}
\end{figure}

The existence of pairs of neighboring small particles
yields a sharp peak at distances $\sigma_{\rm DD} \lesssim r \lesssim 2.5\sigma_{\rm DD}$ 
for the radial distribution function for D-C ($g_{\rm DC}(r)$)
and D-D ($g_{\rm DD}(r)$) pairs, as shown in Fig. \ref{fig12}. 
Hence, even for so low populations of small
particles, it may be expected that competition between bulk-like dynamics 
and confinement occurs for a signifficant fraction of particles. In order to test 
this hypothesis we have calculated mean squared displacements and density-density correlators
for the subset of D-particles which {\it initially} (i.e., at the time origin for the calculation
of both quantities) have at least one neighboring small (C or D) particle
within a distance $r < 1.3\sigma_{\rm DD}$. This distance approximately corresponds to the  
location of the maximum in $g_{\rm DC}(r)$ and $g_{\rm DD}(r)$.
Fig. \ref{fig13} shows $\langle (\Delta r_{\rm D})^2 \rangle$ and $F_{\rm DD}(q,t)$,
computed for the former subset of D-particles, and compared with the corresponding
quantities computed for {\it all} the D-particles (previously shown in Figs. \ref{fig4}a
and \ref{fig10}a). Relaxation of D-particles having neighboring small particles is,
at intermediate times, faster than the average over all D-particles, which is approached
only for very long times corresponding to the onset of the linear diffusive regime.

Differently from the average over all D-particles, the subset initially 
having neighboring small particles does exhibit anomalous relaxation features.
Mean squared displacements display sublinear behavior over two time decades. The corresponding
exponent decreases with decreasing temperature (Fig. \ref{fig13}a).  
Density-density correlators show a concave-to-convex crossover (Fig. \ref{fig13}b).
Logarithmic relaxation is observed for wavevectors $q \sim 2.5\sigma_{\rm DD}^{-1}$ (Fig. \ref{fig13}a), 
i.e., for length scales $\sim 2.5\sigma_{\rm DD}$
which probe the first minimum in $g_{\rm DC}(r)$ and $g_{\rm DD}(r)$.

We have also computed $\langle (\Delta r_{\rm D})^2 \rangle$ and $F_{\rm DD}(q,t)$
for the subset of D-particles which {\it initially} do not have any neighboring small particle 
within a distance $r < 4.5\sigma_{\rm DD}$, i.e., for initially isolated D-particles. 
Results for this subset (not shown) are hardly distinguishable from those averaged over all D-particles
and reported in Figs. \ref{fig4}a and \ref{fig10}a. 
Hence, initially isolated D-particles do not show anomalous relaxation features as
sublinear behavior for mean squared displacements or concave-to-convex and logarithmic decay
for density-density correlators.

\begin{center}
\bf{d) Very large size disparity}
\end{center}

\begin{figure}
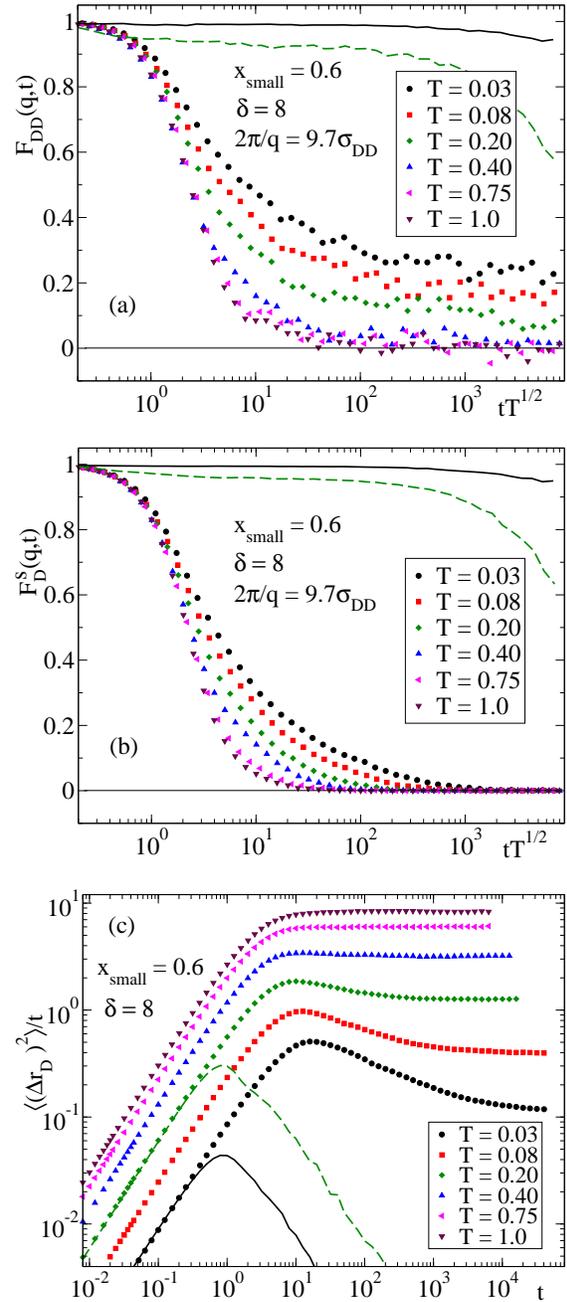

\includegraphics[width=0.895\linewidth]{fig14a.eps}
\newline
\newline
\includegraphics[width=0.895\linewidth]{fig14b.eps}
\newline
\newline
\includegraphics[width=0.895\linewidth]{fig14c.eps}
\newline
\caption{(color online) Simulation results at composition
$x_{\rm small}=0.6$ and size disparity $\delta = 8.0$.
Symbols in panels (a) and (b) are data for, respectively,  $F_{\rm DD}(q,t)$ and $F^{\rm s}_{\rm D}(q,t)$.
Symbols in panel (c) are data for the mean squared displacement of D-particles divided by the time $t$.
Thick solid ($T=0.03$) and dashed ($T=0.20$) curves in the three panels are the corresponding
results for A-particles. Horizontal straight lines in panels (a) and (b) indicate the zero value.}
\label{fig14}
\end{figure}

In this subsection we show and discuss results for the small particles, for a very large size disparity
($\delta = 8.0$) at a single composition $x_{\rm small}=0.6$. Hence, we select a composition for which
small particles exhibit anomalous relaxation features at disparity $\delta = 2.5$ 
(Figs. \ref{fig4}b, \ref{fig6}b and \ref{fig9}). We also use the same packing fraction, $\phi = 0.53$,
that for the case $\delta = 2.5$. We observe a relaxation scenario
rather different from that obtained for disparity $\delta = 2.5$. Fig. \ref{fig14}a
displays density-density correlators for A-A and D-D pairs, at a wavelength
of about 10 times the size of the small particles. In order to provide
a clearer visualization of the decay for the latters, data are plotted as a function
of $tT^{1/2}$, i.e., rescaling the time by the thermal velocity.
As expected, big particles exhibit a much slower dynamics.
Correlators for small particles do not show a slow decay at any temperature. 
Decorrelation occurs in an essentially exponential way down to very small values of $F_{\rm DD}(q,t)$, 
where a tail of small amplitude arises and extends to very long times. 

Below some given temperature a background clearly distinguishable from zero arises. 
The amplitude of the background increases as the system is cooled down. 
Qualitatively similar results (not shown) are observed for other wavevectors.
We identify the background amplitude as an operational non-ergodicity parameter that increases 
from some critical value $f_q^{\rm c}$.
By visual inspection of Fig. \ref{fig14}a it is difficult to establish wether $f_q^{\rm c}$ defined
in this way is zero, but it is evident from data at $T = 0.40$ that  it takes, as much, a extremely small value.
This behavior resembles features characteristic of MCT transitions of the type-A, which are defined
by a zero value of the critical non-ergodicity parameter, in contrast to the finite value defining
standard type-B transitions. 
A recent realization of this scenario has been reported for a system of dumbbell molecules
\cite{chong,dumbbellsprl,dumbbellsjcp}.
While for moderate molecular elongations a standard relaxation scenario is observed, 
for small elongations angular correlators exhibit features analogous to those
of Fig. \ref{fig14}a \cite{dumbbellsjcp}. Theoretical calculations 
for this latter system \cite{chong} relate such features to the existence 
of a MCT transition of the type-A.

Fig. \ref{fig14}b displays self-correlators, 
$F^{\rm s}_{\alpha}(q,t) = \Sigma_{j}\exp\{i{\bf q}\cdot [{\bf r}_{\alpha,j}(t)-{\bf r}_{\alpha,j}(0)]\}$,
for A- and D-particles. As observed for $F_{\rm AA}(q,t)$, self-correlations for A-particles 
show a much slower dynamics. Self-correlators for D-particles
display a striking result. As for density-density correlations (Fig. \ref{fig14}a),
they exhibit a fast decay followed by a long tail. However, they decay to zero at all 
the investigated temperatures, even at those where $F_{\rm DD}(q,t)$ show a finite
long-time limit. Fig \ref{fig14}c shows results for the mean squared displacement.
Data are divided by $t$ in order to evidence that D-particles reach the
diffusive regime (horizontal limit in this representation) at long times.
Data in Figs. \ref{fig14}b and \ref{fig14}c show that
the self-motion of small particles is ergodic at all the investigated temperatures.
A small particle can reach regions arbitrarely far from its initial position.
However, according to results in Fig. \ref{fig14}a, coherent dynamics are non ergodic below some given
temperature. Hence, two small particles can explore the whole structure of the matrix of big particles, 
but they will never fully decorrelate from each other, i.e., the distance between them will always be finite.

These features are consistent with early MCT theoretical calculations by Bosse and Thakur \cite{bosse1,bosse2}
for a binary mixture of hard spheres. Though in that work no information is given
about relaxation features, calculations are reported for non-ergodicity parameters of self- ($f^{\rm s}_q$) 
and density-density ($f_q$) correlators for the small particles. For a symmetric composition and 
large disparity size ($\delta \sim 10$) it is found \cite{bosse1} that $f^{\rm s}_q = 0$ 
at packing fractions where $f_q > 0$, in agreement with results reported here with temperature
as the control parameter.
 
\begin{center}
\bf{e) A possible unified picture}
\end{center}

The whole picture here reported for the small particles at large ($\delta = 2.5$) and very large ($\delta = 8.0$)
size disparity provides a connection with MCT theoretical results by Krakoviack
in Refs. \cite{krakoviackprl,krakoviackjpcm} for a mixture of {\it fixed and mobile} hard spheres. 
In that work the dynamic phase diagram displays an
A- and a B-line in the plane $\phi_{\rm fix}$-$\phi_{\rm mob}$, where $\phi_{\rm fix}$
and $\phi_{\rm mob}$ are the packing fractions of, respectively, the fixed and mobile particles.
The A- and B-lines merge at a higher-order point (namely an A$_3$-point).
The B-line extends from the A$_3$-point to the limit $\phi_{\rm fix} = 0$, where the liquid
of hard spheres is recovered. The A-line extends from the A$_3$-point to the Lorentz gas
limit at $\phi_{\rm mob} = 0$.

For high concentrations of the mobile particles, 
the matrix of fixed particles does not yield signifficant confinement effects. The only
caging mechanism is normal hard-sphere repulsion at short length and time scales,
and dynamic correlators exhibit a standard two-step decay \cite{krakoviackjpcm}. The transition
point is of the type-B and hence the jump of the long-time limit of the density-density correlator
is finite \cite{krakoviackjpcm}, providing a non-zero value of the critical non-ergodicity parameter. 

For high dilution of the mobile particles, hard-sphere repulsion at short length and time scales
cannot yield temporary caging, and density-density correlators display a fast decay
to values close to zero \cite{krakoviackjpcm}. At longer time scales the mobile particles 
probe the structure of the confining matrix of fixed particles, which leads 
to a ``mesoscopic'' caging, characterized by a length scale
larger than that characteristic of bulk-like hard-sphere repulsion. As a consequence of this large-scale
caging mechanism (confinement), density-density correlators for the mobile particles exhibit a long tail of small
amplitude after the fast microscopic decay \cite{krakoviackjpcm}. At the transition point, of the type-A,
the long-time limit does not exhibit a finite jump but grows up continuously \cite{krakoviackjpcm},
providing a zero value for the critical non-ergodicity parameter.

As mentioned above, at moderate concentrations of fixed and mobile particles,
a higher-order A$_3$-point arises \cite{krakoviackprl}
as a consequence of the competition between the mentioned ``microscopic'' and ``mesoscopic'' caging
mechanisms. Relaxation features have not been reported in Refs. \cite{krakoviackprl,krakoviackjpcm}
for state points close to the A$_3$-point. However, as stressed in \cite{krakoviackprl},
they will {\it necessarily} display the anomalies reported here, as a mathematical consequence 
of the value $\lambda =1$ defining the A$_3$-point.

How do results in Refs. \cite{krakoviackprl,krakoviackjpcm} compare with relaxation features presented here?.
First it is worth emphasizing that, differently from Krakoviack's work, high dilution of the small particles
is not the key ingredient for yielding a type-A transition for the latters in the present system.
Indeed, data reported here for $\delta = 2.5$ and $x_{\rm small}=0.1$ show features rather different
from those characterizing type-A transitions. It must be noted that, for these control parameters,
the system here investigated is much denser than at state points close to the A-line for the mixture of fixed 
and mobile particles \cite{krakoviackprl}. Despite its low density, diffusion of the mobile
particles in the latter system is blocked {\it at large length scales} due to the absence
of percolating free volume. In the system here investigated, small particles can diffuse 
at higher densities due to the non-static nature of the confining matrix. 
The slow motion of the big particles creates regions of sufficient local free volume 
which facilitate diffusion of the small particles. As a consequence of high density,
short-range bulk-like caging is a relevant arrest mechanism for small particles and leads to a slow 
decay of dynamic correlators (Fig. \ref{fig10}a), different from features characteristic of type-A transitions.
Hence, for disparity $\delta = 2.5$, type-A transitions cannot exist at any mixture composition.

\begin{figure}
\includegraphics[width=0.85\linewidth]{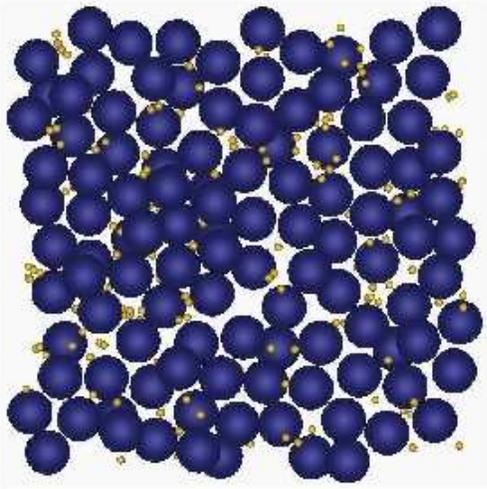}
\newline
\caption{(color online) Typical slab (of thickness $10\sigma_{\rm DD}$) 
at $T=0.03$ and $x_{\rm small}= 0.6$ for size disparity $\delta = 8.0$.
Dark: A- and B-particles. Light: C- and D-particles.}
\label{fig15}
\end{figure}

A way for yielding a type-A scenario for the small particles in the present system,
at a fixed packing fraction and mixture composition, is by increasing the size disparity, as shown above.
In this situation small particles move in a medium of low local density, and short-range 
bulk-like caging is supressed. This effect is illustrated in Fig. \ref{fig15} by displaying a typical slab
of the simulation box. However, small particles are blocked at large length scales below
some given temperature where the tail of the soft-sphere potential is probed.
At that temperature, despite of local vibrations of the confining matrix,
the effective radius of the big particles becomes so large
that the percolating free volume vanishes, leading to a dynamic picture analogous to the
type-A scenario observed for the system of Refs. \cite{krakoviackprl,krakoviackjpcm}.

Relaxation features observed for the small particles at $\delta = 8.0$ and $x_{\rm small}=0.6$
suggest the presence of a nearby A-line, originating from large-scale caging (confinement)
induced by the slow matrix of big particles. For small size disparity, $\delta \rightarrow 1$, 
one obviously recovers the standard MCT scenario
predicted for the monoatomic liquid of hard spheres and widely observed in mixtures of
hard \cite{foffipre}, soft \cite{roux}, or Lennard-Jones spheres \cite{koband}, where small disparity 
is introduced just to avoid crystallization. The standard scenario, originating
from short-range bulk-like caging, is characterized by a nearby B-line. 

Increasing the size disparity at fixed composition $x_{\rm small}=0.6$ will weaken
the effects of bulk-like caging and strengthen those associated to confinement. This feature suggests 
that a crossover from a B- to an A-line will occur by increasing the value of $\delta$,
in analogy with results for the system investigated in Refs. \cite{krakoviackprl,krakoviackjpcm},
where such a crossover is obtained by varying the mixture composition.
For the latter system, the B- and A-lines merge at a higher-order point.
The resemblance of results reported here for $\delta = 2.5$ and $x_{\rm small}=0.6$
(Figs. \ref{fig4}b, \ref{fig6}b and \ref{fig9}) with relaxation features 
characterizing higher-order MCT transitions
suggest an analogous merging. The former results would originate from the presence 
of a nearby B-line (due to the finite value of the non-ergodicity parameters), 
ending at a nearby higher-order point (which would produce the observed anomalous relaxation features),
or at least at an A$_2$-point with $\lambda$ very close to unity.

\begin{center}
{\bf VI. CONCLUSIONS}
\end{center}

We have carried out simulations on a mixture of big and small particles.
Slow dynamics have been investigated for a broad range of temperature and mixture composition.
The introduction of a signifficant size disparity yields very different time scales
for big and small particles, reaching differences of 2-3 decades in diffusivity
for the lowest investigated temperatures. This model exhibits non-conventional 
relaxation features. Mean squared displacements display sublinear behavior
at intermediate times. The exponent for the corresponding power law decreases by decreasing
temperature. By varying temperature or wavevector, a concave-to-convex crossover is obtained
for the shape of the decay of density-density correlators. At some intermediate point of
this crossover, the decay is purely logarithmic.

These anomalous relaxation features, which are observed over time intervals 
extending up to four decades, strongly resemble predictions of the Mode Coupling Theory (MCT)
for state points close to higher-order transitions, which originate from the
competition between different mechanisms for dynamic arrest. 
By varying the mixture composition, anomalous relaxation is displayed by both 
the big and the small particles. For the big particles we suggest 
competition between soft-sphere repulsion
and depletion effects induced by neighboring small particles. For the small particles we suggest 
competition between bulk-like dynamics and confinement, respectively induced
by neighboring small particles and by the slow matrix of big particles. 

We have also performed simulations at a fixed composition for a very large size disparity.
A new relaxation scenario arises for the small particles,
showing features characteristic of nearby MCT transitions of the type-A. 
This feature provides a connection with MCT theoretical results for a mixture
of mobile and static particles, which report a dynamic phase diagram displaying an A- and a B-line
merging at a higher-order point. A similar crossover is suggested for the system here investigated
by varying the size disparity.

Simulation results reported here do not constitute a rigorous proof of an inherent MCT scenario
as the one described above. A proper answer to this question can only be provided by solving the
corresponding MCT equations. However, the highly non-trivial observed analogies
suggest to consider it as a plausible hypothesis. 
Though being beyond the scope of this article, the observation of confinement effects 
also suggests to test a more speculative hypothesis:
a MCT scenario for a liquid of spheres in fractal dimensions
as the origin of the observed relaxation features for the small particles.
To our knowledge no MCT calculations of mean squared displacements
or dynamic correlators are available for the latter or similar systems.

\begin{center}
\bf{ACKNOWLEDGEMENTS}
\end{center}
   
We thank E. Zaccarelli, F. Sciortino, T. Voigtmann, V. Krakoviack, 
and J. Horbach for useful discussions.
We acknowledge support from the projects NMP3-CT-2004-502235 (SoftComp), 
MAT2004-01017 (Spain), and 206.215-13568/2001 (GV-UPV/EHU Spain).


\begin{thebibliography}{9}
\bibitem{notematrix} Slow dynamics of mobile particles in matrixes of static particles have been
investigated within the MCT framework in several simulation studies \cite{gallo}. In these works 
the relative contribution of the static particles to the total packing fraction is about a 30 \%.
Though quantitive deviations from the bulk counterpart
are obtained for the mobile particles, no evidence of anomalous relaxation features
are observed, and a description in terms of the standard MCT scenario is possible.
According to the theoretical work in Ref. \cite{krakoviackprl}, the higher-order MCT transition
would be located at a relative contribution of the static particles of about a 40 \%.
It is well-known that MCT usually overestimates the location of the transition point.
Presumably, computational tests on the system investigated in Ref. \cite{krakoviackprl}
would provide a higher value for the relative contribution of static particles, and therefore
far from that used in Ref. \cite{gallo}. Though no direct comparison is possible with
the system here investigated (big particles are slower than small ones, not static),
we mention that for the highest value of $x_{\rm small} = 0.8$ here investigated at size disparity
$\delta =2.5$ the relative contribution of big particles to the total packing fraction is a 80 \% .
\bibitem{notecrys} 
We have observed some partial crystallization of the big component 
for the composition $x_{\rm small}=0.1$ at temperature $T \sim 0.70$. No crystallization effects
are observed at this concentration for the other investigated temperatures $T=1.0$ and $T \le 0.50$.
For the latter range, the matrix of big particles, which is quenched from high temperature,
is out-of-equilibrium (see Section III). Partial crystallization would presumably
occur for sufficiently long computional times.
\bibitem{noteorder} A MCT transition of order $n$, A$_n$, is characterized by $n$ solutions
of the critical non-ergodicity parameter, only the highest value being physical.
\bibitem{notefqc} A correction to $f^{\rm c}_q$, of the order of the distance to the transition point,
is neglected in Eq. (\ref{eqlog}).
\bibitem{notelowq} For low-$q$ values, fits in Fig. \ref{fig8}a only reproduce a reduced time window
of the simulation results. However, this is not neccesarily a failure of MCT predictions in this $q$-range. 
A similar example is reported in Ref. \cite{foffipre} for a binary mixture of moderate size disparity.
As explained there, if one is not extremily close to the transition point,
it can accidentaly happen for some $q$-values that the third-order term 
of the series expansion (which is usually neglected
in fitting procedures) is not actually negligible as compared to the second-order term.
Hence a fit to a second-order expansion as Eq. (\ref{eqlog}) would provide a good description of the data
only for a reduced time interval. One could fit data in Fig. \ref{fig8}a by adding a third-order term, 
but the fitting parameters would be strongly coupled and their so-obtained values 
would not be reliable. Anyway, a proper answer to this point can only be provided
by solving the corresponding MCT equations.
\end{thebibliography}
\end{document}